\documentclass[preprint2,times,tighten]{aastex6}
\usepackage{lineno}
\usepackage{color}
\usepackage{enumitem}
\usepackage{hyperref}
\usepackage{verbatim}
\usepackage{amsmath,amstext,mathrsfs}
\usepackage[all]{hypcap} 
\usepackage{afterpage}
\usepackage{graphicx}

\usepackage[utf8]{inputenc}
\usepackage[english]{babel}
\usepackage{natbib}
\newcommand{\vc}[1]{\textbf{\em #1}}

\begin{document}
\title{Energetic Particle Acceleration in Compressible Magnetohydrodynamic Turbulence}

\author{Jian-Fu Zhang\altaffilmark{1} and Fu-Yuan Xiang\altaffilmark{1}} 
\email{ jfzhang@xtu.edu.cn}
\altaffiltext{1}{Department of Physics, Xiangtan University, Xiangtan, Hunan 411105, China}

\begin{abstract}
Magnetohydrodynamic (MHD) turbulence is an important agent of energetic particle acceleration. Focusing on the compressible properties of magnetic turbulence, we adopt test particle method to study the particle acceleration from Alfv\'en, slow and fast modes in four turbulence regimes that may appear in a realistic astrophysical environment. Our studies show that (1) the second-order Fermi mechanism drives the acceleration of particles in the cascade processes of three modes by particle-turbulence interactions, regardless of whether the shock wave appears; (2) not only can the power spectra of maximum acceleration rates reveal the inertial range of compressible turbulence, but also recover the scaling and energy ratio relationship between the modes; (3) fast mode dominates the acceleration of particles, especially in the case of super-Alfv\'enic and supersonic turbulence, slow mode dominates the acceleration for sub-Alfv\'enic turbulence in the very high energy range, and the acceleration of Alfv\'en mode is significant at the early stage of the acceleration; (4) particle acceleration from three modes results in a power-law distribution in the certain range of evolution time. From the perspective of particle-wave mode interaction, this paper promotes the understanding for both the properties of turbulence and the behavior of particle acceleration, which will help insight into astrophysical processes involved in MHD turbulence.
\end{abstract}


\section{Introduction}\label{intro}
The visibility of cosmic astrophysics is due to the presence of a non-thermal particle population that dominates emission spanning more than 20 orders of magnitude of the electromagnetic spectrum (e.g., \citealt{Longair11}). Correctly explaining the essence of astrophysical activity phenomenon is to understand how non-thermal high-energy particles are accelerated. The widely accepted effective mechanisms for the acceleration of high-energy particles are the diffusion shock wave (termed as the first-order Fermi process; \citealt{Bell78}) and the turbulent magnetic reconnection (\citealt{LV99}, hereafter LV99; \citealt{Lazarian15,Lazarian20} for recent reviews), the latter of which has been also considered to be a first-order Fermi acceleration process (\citealt{degouveia05,Kowal12,Zhang21}). At present, the second-order Fermi acceleration has been suggested for acceleration of particles in a number of astrophysical settings, such as black-hole X-ray binary (\citealt{Zhang18b}), supernova remnant (\citealt{Fan10}), solar physics (\citealt{Yan08,Petrosian08}), the interstellar medium (\citealt{Cho03}) and diffuse intracluster medium in galaxy clusters (\citealt{Brunetti16}). It is noticed that MHD turbulence is an important agent of particle acceleration as pointed out by \cite{Alfven42} for stochastic particle acceleration.

The new paradigm of MHD turbulence, especially compressible properties associated with Alfv\'en, slow and fast modes, has significantly promoted the understanding of energetic particle-turbulence interactions, including particle acceleration, scattering, diffusion and propagation processes (\citealt{Yan02,Yan04,Chand03,Cho06,Beresnyak16,Xu13,Xu18,Lazarian21}). For instance, stochastic particle acceleration from slow mode (\citealt{Chand03}) and fast one (\citealt{Yan08}) assuming efficient pitch-angle scattering was applied to elucidating solar flares.  \cite{Cho06} made use of the established scaling of slow and fast modes to provide a systematic analytical research of particle acceleration in strong and weak MHD turbulence. In the limits of both slow and fast particle diffusion, they found that fast mode accelerates particles more efficiently than slow mode, the efficiency of particle acceleration by slow and fast modes depends on the spatial diffusion rate, and highly supersonic turbulence is an efficient agent for the acceleration. Since Alfv\'en and slow modes undergo an anisotropy cascade mainly in the direction perpendicular to the underlying mean magnetic field, they are generally considered the comparatively inefficiency in the acceleration of particles. This needs further verification of numerical simulation.

Generally, test particle method is popular in studying the interaction between cosmic ray particles and MHD turbulence. In this way, to trace particle trajectories by integrating the Lorentz force, \cite{Xu13} pointed out that cosmic ray diffusion coefficients are spatially correlated to the local magnetic turbulence properties and confirmed that cosmic rays are superdiffusive in MHD turbulence. Similarly, the recent study using the open source {\it PENCIL} code demonstrated that the fast mode dominates the cosmic ray propagation, whereas Alfv\'en and slow modes are much less efficient with similar pitch angle scattering rates (\citealt{Maiti21}). With the purpose of studying the perpendicular and parallel diffusion of magnetic field, they both ignored the effect of the induced electric field $\vc E= -\vc v \times \vc B$ arising from MHD turbulence. When investigating particle acceleration in a turbulent reconnection layer or a purely turbulent setting, kinetic equation including the induced electric field has been numerically solved using the similar test particle method (\citealt{Kowal11,Kowal12,delV16,Zhang21}).       

Recently, a variety of observation-based new techniques has shed new light on the understanding of the nature of MHD turbulence (e.g., \citealt{LP12, Zhang16,Zhang18,Zhang20}). Especially with application of the quadrupole ratio modulus for synchrotron polarization intensity, \cite{Wang20} successfully recovered the anisotropic properties of compressible MHD turbulence. Statistics of the polarization intensity from Alfvén and slow modes demonstrate the significant anisotropy, while the statistics of the polarization intensity from fast mode show isotropic feature. This is in agreement with earlier directly numerical results provided in \cite{Cho02}. These findings indeed show that synchrotron polarization radiation signature is closely related to MHD turbulence properties. In almost all the studies that we know currently, the specific acceleration properties of non-thermal high-energy particles have still not been explored when applying techniques to the measurement of the turbulence. The purpose of our current work is to explore how high-energy particles are accelerated in compressible MHD turbulence, and whether the acceleration of high-energy particles are relevant to the properties of compressible MHD turbulence. 

We organize the main structure of this paper  as follows: theoretical description of MHD turbulence and particle acceleration mechanism in Section \ref{Theor}, simulation methods in Section \ref{Smethod}, simulation results in Section \ref{Sresult}, discussions in Section \ref{Dis} and summary in Section \ref{Sum}. 

\section{Theoretical Description}\label{Theor}
\subsection{MHD turbulence theory}
It is widely accepted that Goldreich \& Sridhar (1995, hereafter GS95) initiated the development of modern MHD turbulence theory (see \citealt{Beresnyak19b} for a recent review). In the case of strong turbulence, that is, $M_{\rm A}=V_{\rm L}/V_{\rm A}\simeq 1$, where $V_{\rm L}$ is the injection velocity at the scale $L_{\rm in}$ and $V_{\rm A}=B_0/\sqrt{4\pi \rho}$ is the Alfv\'enic velocity with the mean magnetic field strength $B_0$ and the density $\rho$, this work predicted the scale-dependent anisotropy ($l_{\|}\propto l_{\perp}^{2/3}$) of incompressible MHD turbulence through the assumption of critical equilibrium, i.e., $v_{l}l_{\perp}^{-1}=V_{\rm A}l_{\|}^{-1}$. Here, $v_ l$ indicates the fluctuation velocity of turbulence at the scale $l$, while $l_{\|}$ and $l_{\perp}$ denote parallel and perpendicular scales of the eddies, respectively.    
 
In the framework of a collection of anisotropic eddies for MHD turbulence, GS95 was generalized to both $M_{\rm A}<1$ and $M_{\rm A}>1$ (LV99 and \citealt{Lazarian06}). In the case of $M_{\rm A}<1$, i.e., the MHD turbulence being driven with sub-Alfv{\'e}nic velocities at injection scale $L_{\rm in}$, it was found that a weak turbulence spans from $L_{\rm in}$ to the transition scale $l_{\rm tr}=L_{\rm in}M_{\rm A}^2$, while the strong sub-Alfv{\'e}nic turbulence is in the range from the dissipation scale $l_{\rm dis}$ to the transition scale $l_{\rm tr}$, where the turbulence velocity $v_l$ is equal to $V_{\rm A}$. In this strong turbulence range, we have the relation 
\begin{equation}
l_{\|}\approx L_{\rm in}^{1/3}l_{\perp}^{2/3}M_{\rm A}^{-4/3} \label{anis}
\end{equation}
between the parallel scale $l_\|$ of the eddy extended along the magnetic field and its transversal scale $l_\bot$. When setting $M_A=1$, we obtain the original GS95 relation of $l_{\|}\propto l_{\perp}^{2/3}$. 

The opposite case, $M_{\rm A}>1$, corresponds to super-Alfv{\'e}nic turbulence, i.e., $ V_{\rm L} > V_{\rm A}$, where $V_{\rm L}$ is the turbulence injection velocity. For a limiting case of ${M}_{{\rm{A}}}\gg 1$, since the weakly turbulent magnetic field undergoes a marginal influence of MHD turbulence, the turbulence at the scale close to $L_{\rm in}$ scale has an essentially hydrodynamic Kolmogorov performance, i.e., $v_l=V_{\rm L}(l/L_{\rm in})^{1/3}$. Interestingly, changes happen at the scale of $l_{\rm A}=L_{\rm in}M_{\rm A}^{-3}$  to the hydrodynamic property of turbulence cascade (see Section 2.2 in \citealt{Lazarian06}), where the turbulent velocity is equal to the Alfv{\'e}n velocity. In the inertial range from $l_{\rm A}$ to  $l_{\rm dis}$, the super-Alfv{\'e}nic turbulence follows again the characteristics of the GS95 anisotropy (GS95 and \citealt{Lazarian06}).
 
 Another important component of MHD turbulence theory is its compressibility (\citealt{Beresnyak19} for recent review; \citealt{Beresnyak19b} for textbook), which was confirmed in numerical simulations (\citealt{Lith01,Cho02,Cho03} ) using the Fourier (\citealt{Cho02}) or wavelet (\citealt{Kowal10}) decomposition MHD turbulence into three modes, i.e., Alfv\'en, slow and fast modes. It was demonstrated that while density is seriously modified by compressibility, the magnetic and velocity fluctuations of Alfv{\'e}n and slow modes are only slightly different from that of the incompressible case (\citealt{Cho02,Kowal10}). Generally, Alfv{\'e}n modes are compatible with the GS95 anisotropic property ($l_{\|} \propto l_{\perp}^{2/3}$), following Kolmogorov spectrum of $E\propto k^{-5/3}$, and slow modes sheared by Alfv\'en modes are evolved passively, showing the GS95 anisotropic property, while fast modes have isotropic properties ($l_{\|} \propto l_{\perp}$), abiding by the scaling of $E\propto k^{-3/2}$ (\citealt{Cho02}). What is worth noting is that the properties of these modes have been well confirmed by observational research, such as synchrotron radiation statistics (\citealt{LP12,Wang20}) and velocity channel analysis of spectroscopic measurement (\citealt{Kandel16,Kandel17}). 
 
 \subsection{Particle acceleration mechanism}
Particle acceleration mechanism can generally be classified as dynamic, hydrodynamic and electromagnetic processes (\citealt{Longair11}). In general, magnetic fields themselves do no work and cannot be directly responsible for particle acceleration, but time-varying magnetic fields lead to an inductive electric field, $\vc{E}=-\partial \vc{B}/ \partial t$, to accelerate particles. The classic model of particle acceleration dates back to \cite{Fermi49} in which particles can statistically gain energy through collisions with interstellar clouds, resulting in a second-order acceleration, $\Delta E/E\propto (V/c)^2\sim O(V^2/c^2)$, where $V$ and $c$ are the velocity of the cloud and the speed of light, respectively. By assuming a characteristic escape time, $t_{\rm esc}$, he obtained a distribution of particle energies, $N\propto E^{-m}$ with $m=1+( \alpha t_{\rm esc})^{-1}$, where $\alpha$ and $t_{\rm esc}$ are model-dependent parameters. Thus, this model does not predict a single power-law exponent, i.e., a change in the power-law index. In the modern understanding of the second-order Fermi acceleration (\citealt{Longair11}), the particles gain energy by interacting with various types of plasma wave or irregularities in the interstellar magnetic field. As for this mechanism, it is generally thought that there are few chances of a particle gaining significant energy due to the second-order behavior of $O(V^2/c^2)$.
 
 To enhance acceleration efficiency, many researches on particle acceleration have been motivated independently and applied to many diverse astrophysical environments (\citealt{Bell78,Blandford78,Drury99}). The acceleration process was associated with acceleration in strong shock waves, referred to as diffusive shock acceleration, in which the acceleration is first-order, $\Delta E/E\propto (V/c)\sim O(V/c)$ in the shock velocity and predicts a power-law spectrum with an index of 2. This model, as first-order Fermi acceleration, is efficient for accelerating particles. However, this model ignored detailed physics of the magnetic field and energy losses of the particles. Considering the effect of magnetic fields, it is involved in a complex non-linear process and causes a more of less universal power-law energy spectrum (\citealt{Berezhko07,Kang06,Kang09}).  

In analogy to classical diffusive shock acceleration, it was suggested that a first-order Fermi process operates within the 3D current sheet of turbulent reconnection (LV99, \citealt{degouveia05,Drury12,Kowal11,Kowal12,Zhang21}), where particles trapped in the converging magnetic fluxes of opposite polarity bounce back and forth in the large-scale reconnection zone. Similarly, \cite{Drake06} proposed that the particles trapped in contracting 2D magnetic islands or plasmoids caused by tearing reconnection also undergo a similar first-order Fermi process. In driving the first-order Fermi process, it was demonstrated to be equivalent between plasma kinetic scale reconnection and MHD turbulent reconnection (\citealt{Kowal11}).

\section{Simulation Methods}\label{Smethod}
\subsection{Simulation of MHD Turbulence }
The equations describing MHD turbulence are given as follows 
\begin{equation}
{\partial \rho }/{\partial t} + \nabla \cdot (\rho {\vc v})=0,
\end{equation}
\begin{equation}
\rho[\partial {\vc v} /{\partial t} + ({\vc v}\cdot \nabla) {\vc v}] +  \nabla p
        - {\vc J} \times {\vc B}/4\pi ={\vc f},
\end{equation}
\begin{equation}
{\partial {\vc B}}/{\partial t} -\nabla \times ({\vc v} \times{\vc B})=0,
\end{equation}
\begin{equation}
\nabla \cdot {\vc B}=0, \label{eq:14}
\end{equation}
where $p=c_{\rm s}^2\rho$ is the gas pressure, $c_{\rm s}$ the sonic speed, $t$ the evolution time of the fluid, ${\vc J}=\nabla \times {\vc B}$ the current density, and ${\vc f}$ a random driving force. In our simulation, we use a third-order-accurate hybrid, essentially non-oscillatory code to solve the above control equations in a periodic box at the length of $2\pi$. In wavenumber space, we drive the turbulence by a random solenoidal driving force on large scale (small wavenumber $k\simeq2.5$ ), and set a non-zero mean magnetic field along the $x$-axis direction. The data cubes with numerical resolution of $512^3$ we obtained, corresponding to four different turbulence regimes, are listed in Table \ref{table:simdata}, where each group of data is mainly depicted by the plasma parameter, the Alfv\'enic and sonic Mach numbers. 

\cite{Cho02}  provided a method to decompose MHD modes into Alfv\'en, slow and fast modes. The main procedure of this decomposition method is characterized by the following displacement vectors
\begin{equation}
\hat{\zeta}_{\rm f} \varpropto (1+\frac{\beta}{2} + \sqrt{D})(k_{\perp} \hat {\vc k}_{\perp}) 
+(-1 + \frac{\beta}{2}+ \sqrt{D})(k_{\parallel} \hat {\vc k}_{\parallel}), \label{eq:15}
\end{equation}
\begin{equation}
\hat{\zeta}_{\rm s} \varpropto (1+\frac{\beta}{2} - \sqrt{D})(k_{\perp} \hat {\vc k}_{\perp}) 
+(-1 + \frac{\beta}{2}- \sqrt{D})(k_{\parallel} \hat {\vc k}_{\parallel}), \label{eq:16}
\end{equation}
\begin{equation}
\hat{\zeta}_{\rm A} \varpropto -\hat{\vc k}_{\perp} \times \hat{\vc k}_{\parallel},
\end{equation}
where $D=(1+\frac{\beta}{2})^2-2 \beta \cos^2\theta$ and $\cos\theta=\hat{\vc k}_{\parallel} \cdot \hat{\vc B_0}$. By projecting magnetic field and velocity into $\hat{\zeta}_{\rm f}$, $\hat{\zeta}_{\rm s} $ and $\hat{\zeta}_{\rm A}$, we can obtain their components for fast, slow and Alfv\'en modes. It is noticed that when each MHD mode is separated by the Fourier transformation, this technique is closely related to the mean magnetic field in the global frame. In the view of depending on the wandering of large scale magnetic field and density inhomogeneities, this technique is only applicable for the case of the sub-Alfv{\'e}nic turbulence with a strong mean magnetic field and small perturbation. Afterward, \cite{Kowal10} proposed an improved procedure to extend the earlier decomposition (\citealt{Cho02}) by introducing a discrete wavelet transformation before the Fourier separation. This improved method relies on the local magnetic field instead of the mean magnetic field. Therefore, it has more significant advantages in decomposing turbulent fluctuations of high amplitude in the case of the sup-Alfv\'enic turbulence.

\begin{table}[t]
 \centering
 \begin{tabular}{c c c ccc}
Models & $B_0$ & $M_{\rm A}$ & $M_{\rm s}$ & $\beta$ & Description\\ \hline \hline
run1 & 1.0 & 0.65 & 0.48 & 3.67 &  Sub-Alfv{\'e}nic \& subsonic \\
run2 & 1.0 & 0.50 & 9.92 & 0.01 &  Sub-Alfv{\'e}nic \& supersonic \\
run3 & 0.1 & 1.72 & 0.45 & 29.12 &  super-Alfv{\'e}nic \& subsonic \\
run3 & 0.1 & 1.76 & 7.02 & 0.13 &  super-Alfv{\'e}nic \& supersonic \\  \hline \hline
\end{tabular}
 \caption {Data cubes with numerical resolution of $512^3$ for different  turbulence regimes. The mean magnetic field $B_0$ in units of code is set along the $x$-axis. $M_{\rm A}$ is the Alfv{\'e}nic Mach number, $M_{\rm s}$ the sonic Mach number, and $\beta=2M_{\rm A}^2/M_{\rm s}^2$ the plasma parameter. }\label{table:simdata}
\end{table}

\subsection{Method of Test Particle}\label{TestMethod}
In the MHD turbulence, the motion of a charged particle satisfies the following kinetic equation 
\begin{equation}
 \frac{d}{d t} \left( \gamma m \vc{u} \right) = q \left( \vc{E} + \vc{u} \times
\vc{B} \right) , \label{eq:ptraj1}
\end{equation}
where $\gamma \equiv 1/\sqrt{1 - u^2 / c^2}$ is the Lorentz factor of relativistic particle, $\vc{u}$ the particle velocity, $m$ the particle mass and $q$ the particle electric charge. In this work, we consider the acceleration process resulting from magnetic filed $\vc{B}$ and its induced electric field $\vc E=-\vc{v}\times \vc{B}$, igoring resistivity effects of electric field. Consequently, Equation (\ref{eq:ptraj1}) can be rewritten as 
\begin{equation}
 \frac{d}{d t} \left( \gamma m \vc{u} \right) = q \left[ \left( \vc{u} - \vc{v}
\right) \times \vc{B} \right] . \label{eq:ptraj2}
\end{equation}
We integrate this equation to trace particle's trajectories using the 8th order embedded Dormand-Prince method (\citealt{Hairer08}) with adaptive time step. Using
cubic interpolation with the discontinuity detector, we can obtain the local values of the plasma velocity $\vc{v}$ and magnetic field $\vc{B}$ at each step of the integration. In practice, test particles, in the form of a thermal distribution with a temperature of $T= 10^6$ K, are injected into a frozen-in-time 3D MHD domain at a given snapshot. We assume that the speed of light is approximately equal to 20 times greater than the injection velocity and the magnetic field strength is $10^{-3}\  \rm G$. 

\section{Simulation Results} \label{Sresult}
On the basis of data cubes from simulations of MHD turbulence (see Table \ref{table:simdata}) and plasma mode decomposition procedure described above, we in this section provide numerical results for Alfv\'en, slow and fast modes using test particle methods. These results include the evolution of particle kinetic energy over time, the mean acceleration time of test particles, the power spectrum of maximum acceleration rate and spectral energy distributions, ranging from four turbulence regimes:  sub-Alfv\'enic and subsonic, sub-Alfv\'enic and supersonic, super-Alfv\'enic and subsonic, and super-Alfv\'enic and supersonic. 

\begin{figure*}[t]
\centering
\includegraphics[width=0.9\textwidth,height=0.3\textheight,bb=100 80 1020 600]{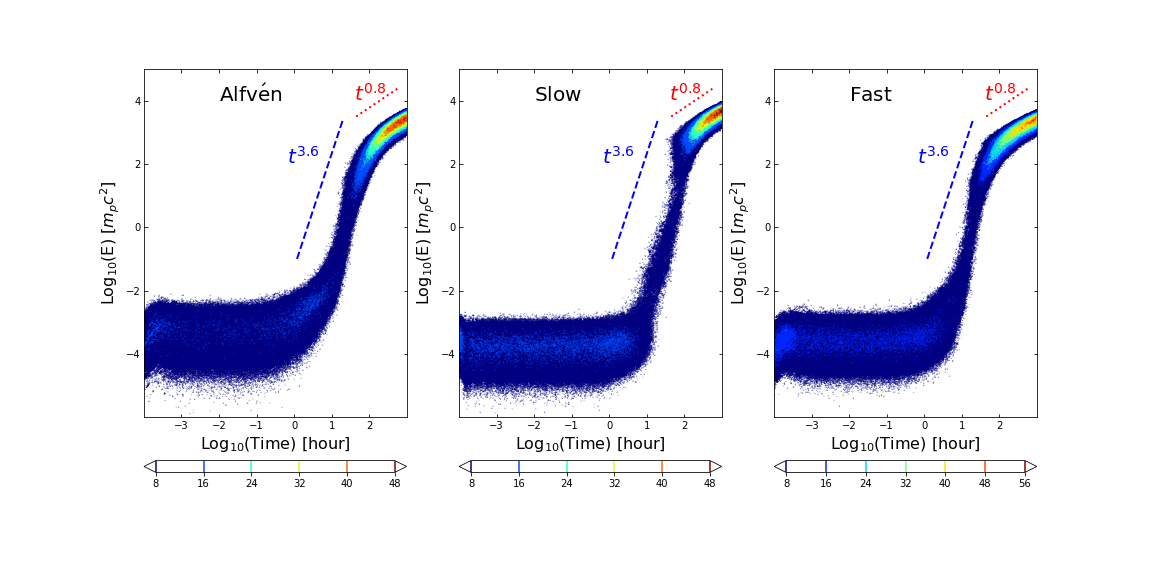} \ \ 
\includegraphics[width=0.9\textwidth,height=0.3\textheight,bb=100 80 1020 580]{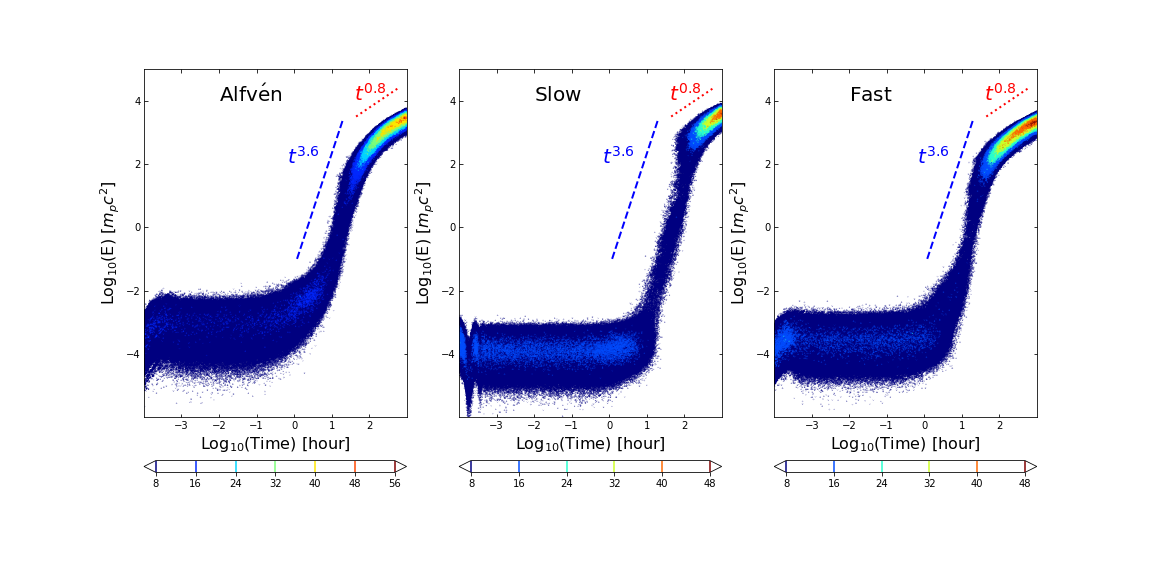} 
\caption{Kinetic energy distributions of 10, 000 protons for Alfv\'en, slow and fast modes, arising from sub-Alfv\'enic \& subsonic (upper panels), and sub-Alfv\'enic \& supersonic (lower panels) turbulence regimes.  In each subpanel, the color bar shows the change in the number of protons.
} \label{fig:ptimesubA}
\end{figure*}

\begin{figure*}[t]
\centering
\includegraphics[width=0.99\textwidth,height=0.3\textheight,bb=100 80 1020 600]{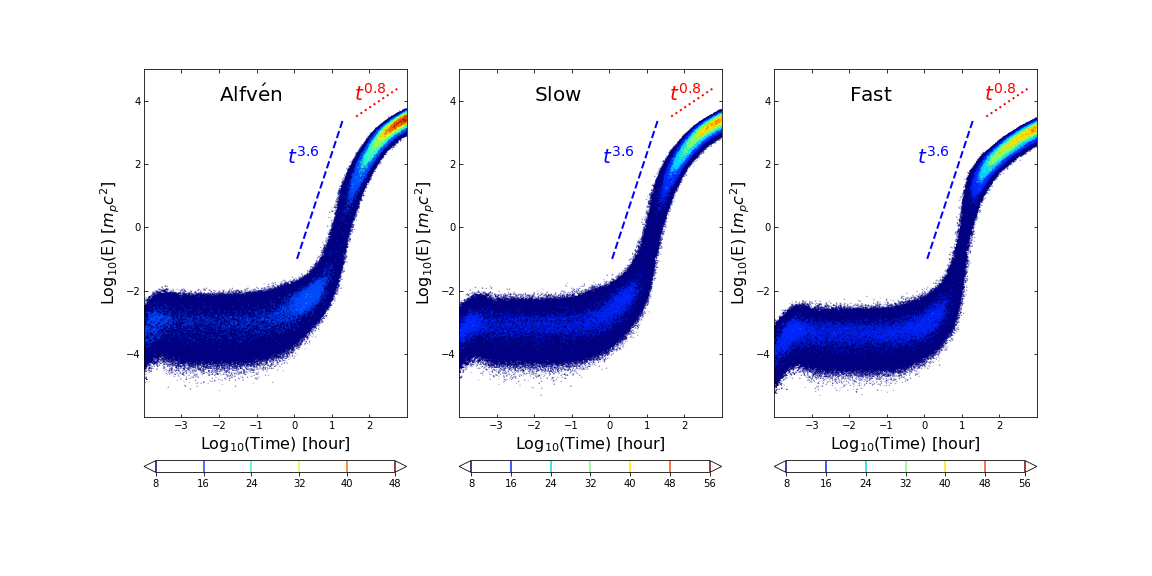} \ \ 
\includegraphics[width=0.99\textwidth,height=0.3\textheight,bb=100 80 1020 580]{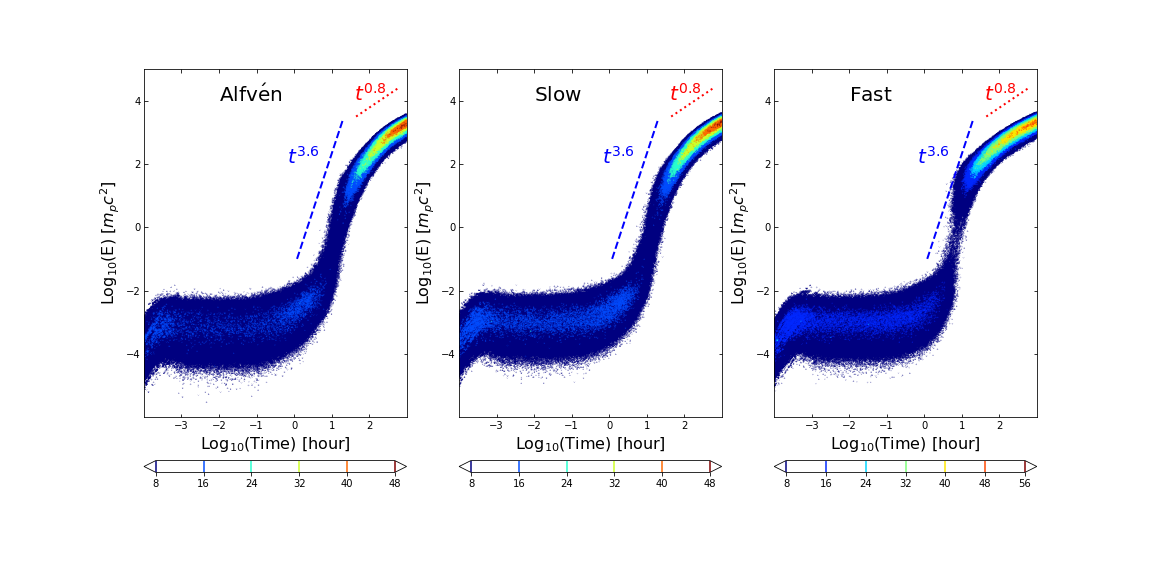} 
\caption{Kinetic energy distributions of 10, 000 protons for Alfv\'en, slow and fast modes, arising from super-Alfv\'enic \& subsonic (upper panels), and super-Alfv\'enic \& supersonic (lower panels) turbulence regimes.  In each subpanel, the color bar shows the change in the number of protons.
} \label{fig:ptimesupA}
\end{figure*}

Figures \ref{fig:ptimesubA} and \ref{fig:ptimesupA} show kinetic energy distributions of 10, 000 protons injected initially in a thermal distribution over the evolution time for Alfv\'en, slow and fast modes, in which various turbulence regimes appearing in astrophysics are explored. Compared to contour maps for three modes, we can see the differences of particle distribution between them, as shown in the color bar in four turbulence regimes. The injected particles first undergo a less efficient acceleration process in the range of less than $t\simeq 5$ hours, and then accelerate significantly with a relation of $E\propto t^{3.6}$. When evolving to the late stage approximately more than $t\simeq 100$ hours, the distribution of particle energy begins to change with the relation of $E\propto t^{0.8}$. Despite the noticeable change of the number of accelerated particle's distribution, it statistically follows a similar distribution relationship under various turbulence conditions from $E\propto t^{3.6}$ to $E\propto t^{0.8}$. 

The less efficient particle acceleration at the early stage should result from the fact that the gyration radius of the test particles is smaller than the eddy size, resulting in insufficient energy exchange from turbulence to particles. Efficient particle acceleration stage, i.e., $E\propto t^{3.6}$, corresponds to full interactions between particles and plasma waves in the inertial range of turbulence. The transition to $E\propto t^{0.8}$ can be understood as a weak turbulence interaction in the spatial region near the injection scale $L_{\rm in}\simeq 200$ code units. Specifically, this corresponds to the scale range of $L_{\rm in}M_{\rm A}^2 < l < L_{\rm in}$ as is in sub-Alfv\'enic turbulence, and $L_{\rm in}M_{\rm A}^{-3} < l < L_{\rm in}$ in super-Alfv\'enic turbulence. In general, this result shows that particle acceleration has universal properties under various turbulent conditions. This implies that the mechanism which causes particles to accelerate is a second-order Fermi acceleration process, comparing with a more effective first-order Fermi acceleration in the turbulent reconnection layer with the relation from $E\propto t^{4.5}$ to $E\propto t^{2.5}$ (\citealt{Zhang21}). Due to the setting of periodic boundary conditions, particles with a gyration radius larger than the box scale will reenter and cause a continuous increase in their kinetic energies, which is similar to a 1.5-order Fermi acceleration in a large-scale turbulence environment around the turbulent reconnection layer (\citealt{Brunetti16,Zhang21}). 

\begin{figure*}[t]   
\centering
\includegraphics[width=0.4\textwidth,height=0.25\textheight,]{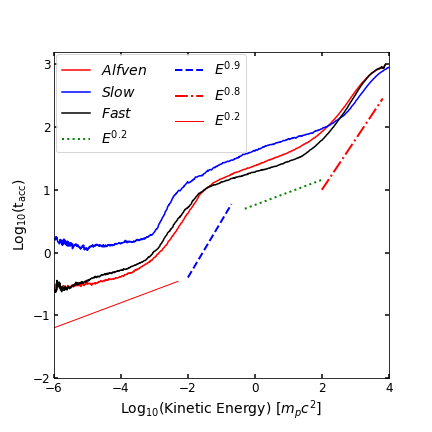} 
\includegraphics[width=0.4\textwidth,height=0.25\textheight,]{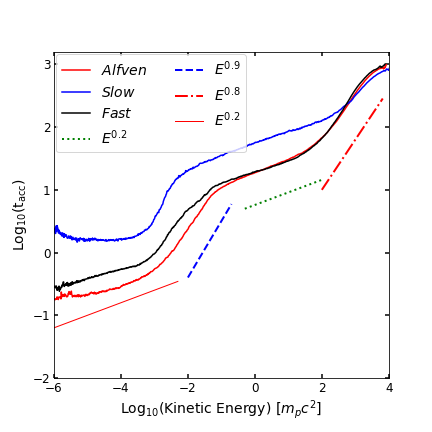} 
\includegraphics[width=0.4\textwidth,height=0.25\textheight,]{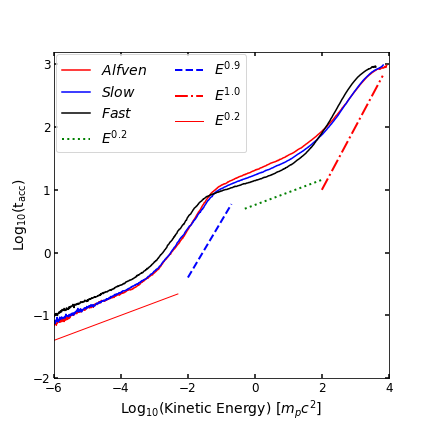} \ \ 
\includegraphics[width=0.4\textwidth,height=0.25\textheight,]{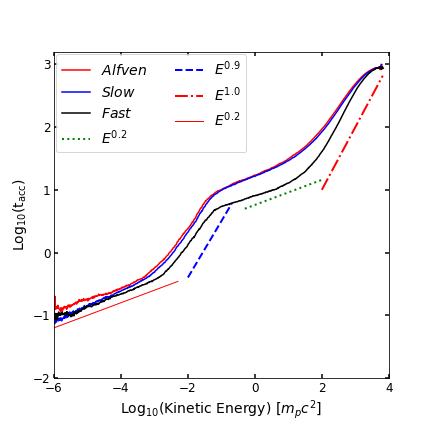} 
\caption{The mean acceleration time as a function of proton kinetic energy for Aflv\'en, slow and fast modes in different turbulence regimes: sub-Alfv\'enic and subsonic (left upper panel), sub-Alfv\'enic and supersonic (right upper), super-Alfv\'enic and subsonic (left lower), and super-Alfv\'enic and supersonic (right lower).
} \label{fig:tacc}
\end{figure*}

In order to further study the acceleration of particles, we adopt the definition of $t_{\rm acc}=\sum_{i=1}^{N_{\rm E}} t_i/N_{\rm E}$ provided in \cite{delV16}, where $t_i$ is the acceleration time of $i$-th particle and $N_{\rm E}$ is the number of particles in the energy interval from $E$ to $E+ \bigtriangleup E$. The mean acceleration time as a function of kinetic energy is plotted in Figure \ref{fig:tacc}, from which we can see more detailed differences between three modes in various turbulence regimes.  As for sub-Alfv\'enic, subsonic (left upper) and supersonic (right upper) regimes, the acceleration of slow mode takes more time than that of Alfv\'en and fast modes in low energy range, both of which accelerate similarly. The mean acceleration time for slow mode decreases with increasing kinetic energy in the very low energy range, but then reverses and increases with energy in the high energy range, representing the transition of power-law relations from $t_{\rm acc}\propto E^{0.9}$ to $t_{\rm acc}\propto E^{0.2}$ then to $t_{\rm acc}\propto E^{0.8}$. Except for the low energy range, it maintains as similar scaling for Alfv\'en and fast modes as that for slow mode. The lower panels corresponding to super-Alfv\'enic, subsonic (left lower) and supersonic (right lower) regimes present that fast mode takes less time than Alfv\'en and slow modes. In particular, there is no remarkable negative correlation between kinetic energy and mean acceleration time in the low energy range. In fact, the power law relationship we find in Figure \ref{fig:tacc} has a one-to-one correspondence to those in Figures \ref{fig:ptimesubA} and \ref{fig:ptimesupA}, that is, $t_{\rm acc}\propto E^{0.9}$ matches with a flat particle distribution in the range of less than about 5 hours, $t_{\rm acc}\propto E^{0.2}$ (plotted in dotted line) with a significantly increasing process of kinetic energy ($E\propto t^{3.6}$) and $t_{\rm acc}\propto E^{0.8}$ with the moderately increasing process at the late stage of evolution ($E\propto t^{0.8}$). 

\begin{figure*}[t]   
\centering
\includegraphics[width=0.46\textwidth,height=0.22\textheight,]{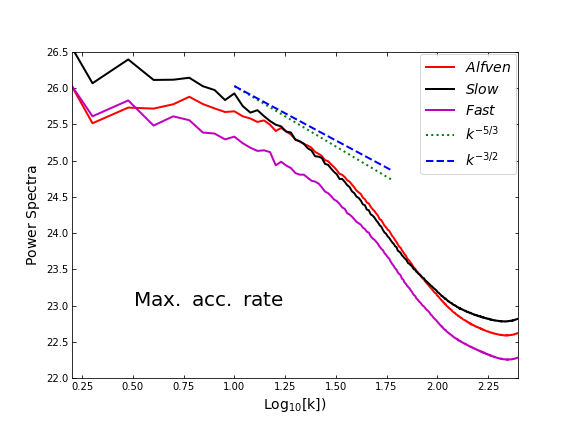} 
\includegraphics[width=0.46\textwidth,height=0.22\textheight,]{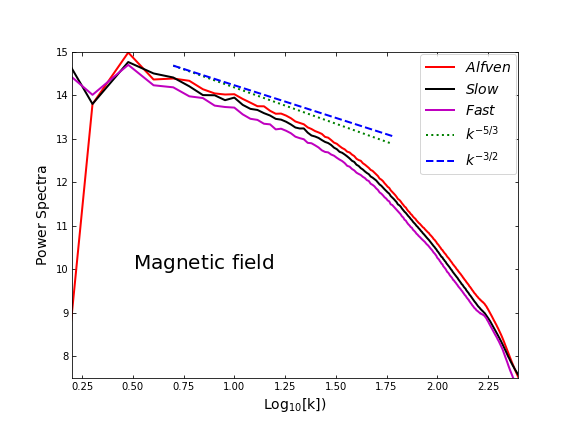} \ \
\includegraphics[width=0.46\textwidth,height=0.22\textheight,]{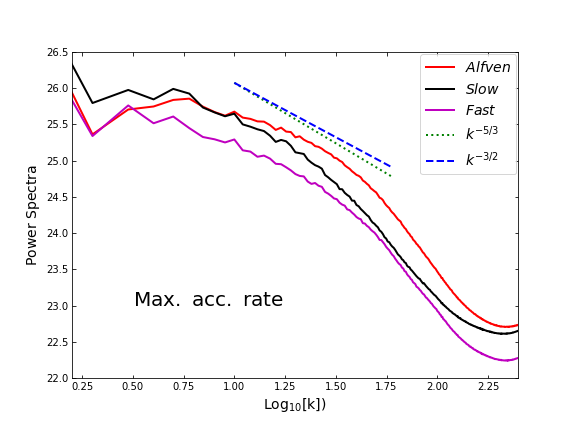} 
\includegraphics[width=0.46\textwidth,height=0.22\textheight,]{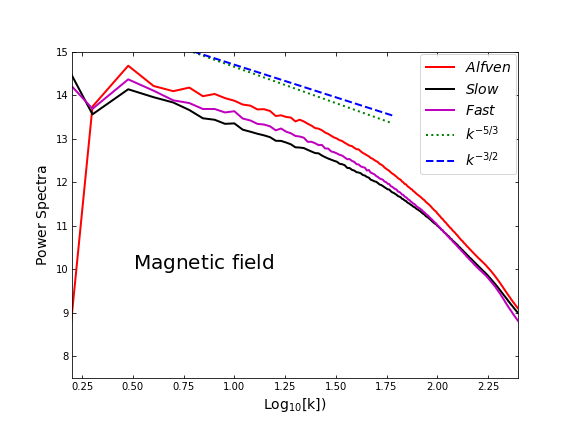}  \ \
\includegraphics[width=0.46\textwidth,height=0.22\textheight,]{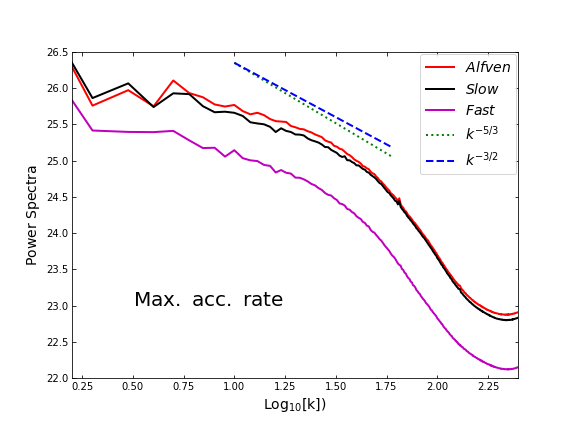} 
\includegraphics[width=0.46\textwidth,height=0.22\textheight,]{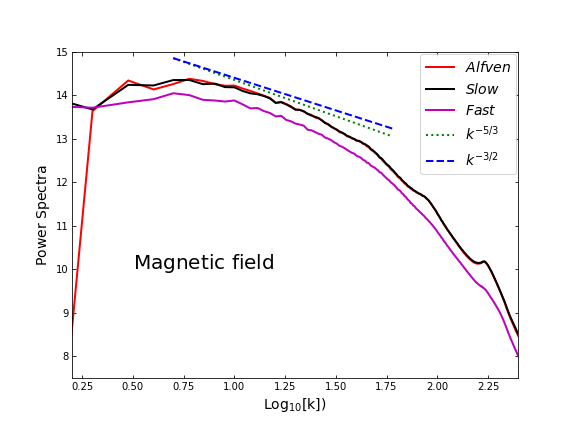}  \ \
\includegraphics[width=0.46\textwidth,height=0.22\textheight,]{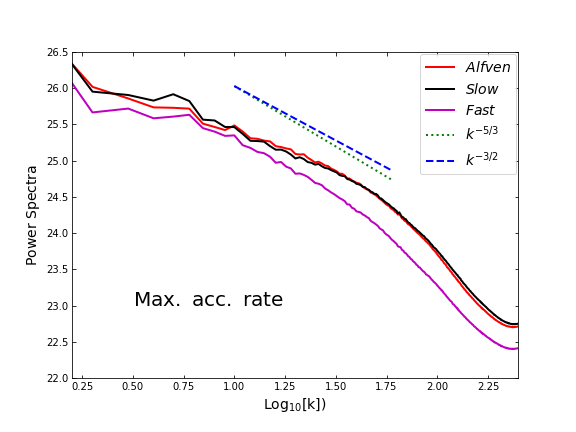} 
\includegraphics[width=0.46\textwidth,height=0.22\textheight,]{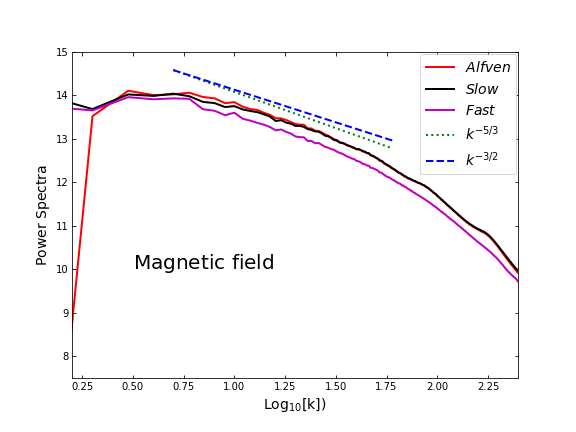} 
\caption{Power spectra of the maximum acceleration rate (left column) and the magnetic field (right column) for Aflv\'en, slow and fast modes in different turbulence regimes:  sub-Alfv\'enic and subsonic (upper panel), sub-Alfv\'enic and supersonic (upper middle), super-Alfv\'enic and subsonic (lower middle), and super-Alfv\'enic and supersonic (lower).
} \label{fig:spec_maxmf}
\end{figure*}

The influence of three modes on the maximum acceleration rate is explored in Figure \ref{fig:spec_maxmf} by analyzing power spectrum, compared to underlying turbulent magnetic fields. The left column of this figure shows power spectra of the maximum acceleration rate of particles, and the right column corresponds to power spectra of the underlying magnetic fields in four turbulence regimes. As shown, the maximum acceleration rates for fast mode present well the scaling of $k^{-3/2}$ in four turbulence regimes explored, while Alfv\'en and slow modes indicate approximately the scaling of $k^{-5/3}$.  Compared to power spectra of magnetic field in the same turbulence regime, we see that there is a more limited inertial range by the maximum acceleration rate, the spectrum of which first becomes steeper at small scales (large wavenumber), then flattens at close dissipation scales. In short, the maximum acceleration rates for three modes reveal the inertial range, the scaling of spectrum and energy exchange dependence from underlying compressible MHD turbulence cascade. Thus, the acceleration efficiency of particles relies on the properties of each mode.

\begin{figure*}[t]
\centering
\includegraphics[width=0.85\textwidth,height=0.22\textheight,bb=80 20 900 435]{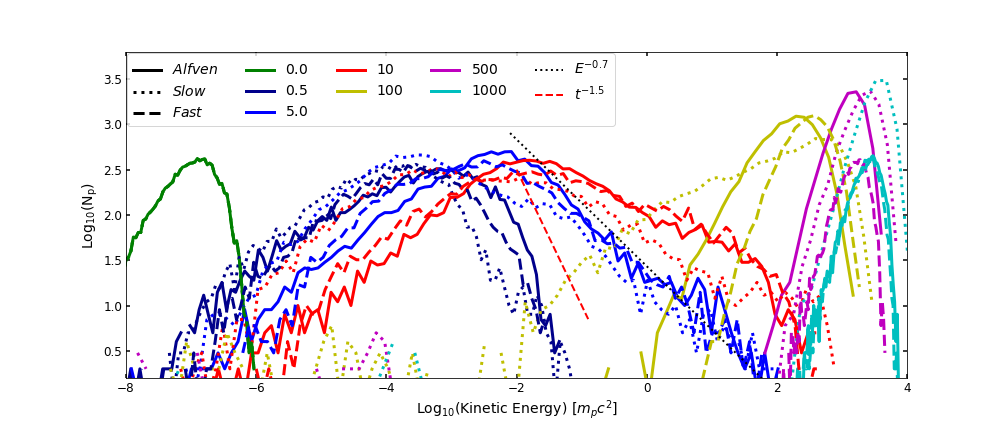} \ \ %,bb=20 20 400 350
\includegraphics[width=0.85\textwidth,height=0.22\textheight,bb=80 20 900 435]{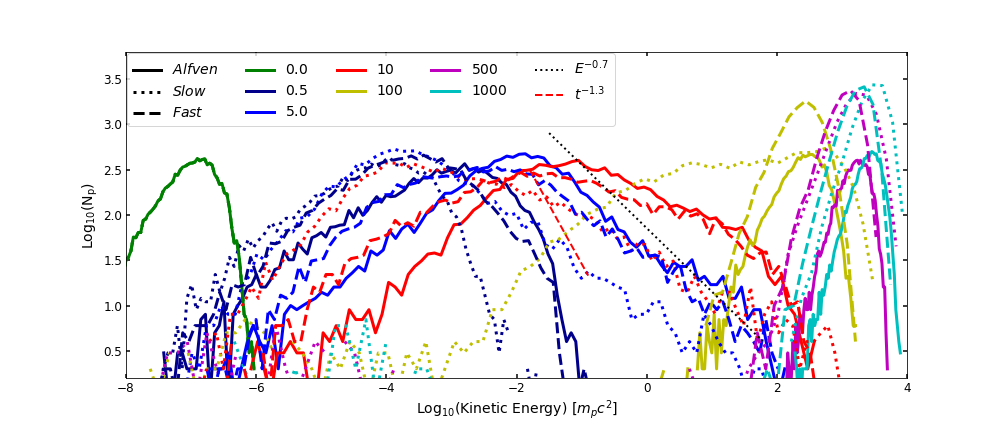} \ \ 
\includegraphics[width=0.85\textwidth,height=0.22\textheight,bb=80 20 900 435]{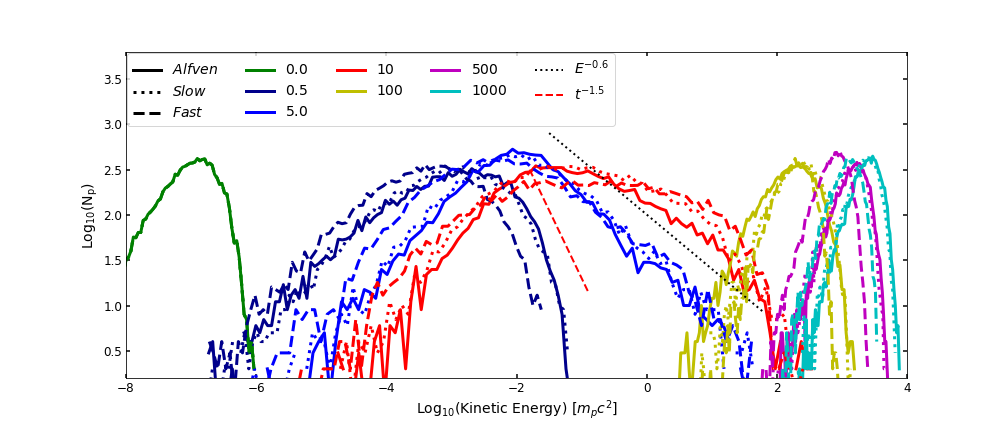} \ \ %,bb=20 20 400 350
\includegraphics[width=0.85\textwidth,height=0.22\textheight,bb=80 20 900 435]{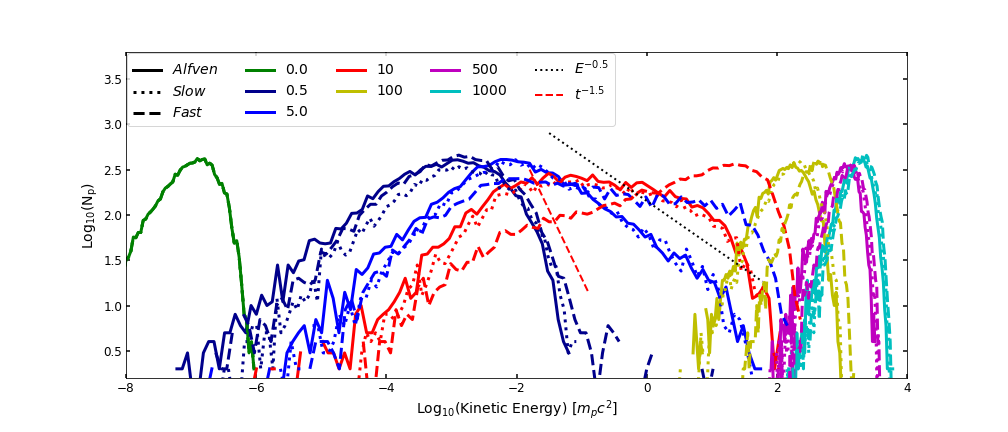} \ \ 
\caption{ Spectral energy distributions of the accelerated protons for Alfv\'en (solid lines), slow (dotted lines) and fast (dashed lines) modes, arising from different turbulence regimes: sub-Alfv\'enic and subsonic (upper panel), sub-Alfv\'enic and supersonic (upper middle), super-Alfv\'enic and subsonic (lower middle), and super-Alfv\'enic and supersonic (lower), at different evolution times in units of hour.
} \label{fig:spec}
\end{figure*}

Furthermore, spectral energy distributions of accelerated particles for three modes are plotted in Figure \ref{fig:spec} at several snapshots $t=0$, 0.5, 5.0, 10, 100, 500 and 1000 hours, in which the thick solid lines indicate spectral distributions for Alfv\'en mode, the thick dotted lines for slow mode, and the thick dashed lines for fast mode.  As time evolves, the particle spectra of the thermal distribution broaden and move to high energy wavebands, and form a changing power-law distribution, e.g., with of $N (E) \propto E^{-1.5}$ at the time $t\simeq5$ hours. In the extremely high energy range, the non-thermal spectra present a thermal-like spectral distribution characteristics. In the case of sub-Alfv\'enic turbulence (upper panel for subsonic, and upper middle panel for supersonic), Alfv\'en and fast modes dominate the acceleration of particles in the range of less than $t\simeq10$ hours, and then it turns to slow mode dominance. In the case of super-Alfv\'enic and subsonic turbulence (lower middle panel), we observe the transition of the plasma modes that dominates the acceleration process, that is, at the early stages of evolution ($t\lesssim 1$ hour), the dominant acceleration process is Alfv\'en and slow modes, then fast mode ($1\lesssim t \lesssim 50$ hours), and finally Alfv\'en and slow modes ($t\gtrsim 50$ hours). In the case of super-Alfv\'enic and supersonic turbulence (lower panel), we find that fast mode dominates the acceleration of particles throughout the entire process. 

In brief, the main features regarding the particle acceleration of three modes are that (1) fast mode plays a major role in accelerating particles in various turbulence regimes we explored; (2) slow mode obviously accelerates particles  in the very high energy part in the case of sub-Alfv\'en turbulence; (3) Alfv\'en mode mainly contributes in the early stage of evolution. 
 
\section{Discussions}\label{Dis}
In the current work we have considered proton as probe particle and injected the number of 10, 000 protons into a frozen-in-time 3D MHD domain that has reached an evolutionary steady state. Due to the fact that the macroscopic MHD dynamical time is much longer than the evolution time of the test particles, it is a reasonable treatment for considering a single snapshot to accelerate the particles. Similar to most PIC acceleration simulations in the collisionless plasma settings, electron-positron pairs can also be regarded as probe particles in the collisional MHD turbulence environment. However, simulating electron-positron pairs evolution in the case of MHD turbulence will be extremely time-consuming because of their more small mean free path, and explored in future work

 We studied the effect of mean magnetic fields on the acceleration process using the structure function analysis of the maximum acceleration rate. We found that the maximum acceleration rate shows an approximate $SF\propto R^{1.4}$ for Alfv\'en, slow and fast modes along three coordinate axes (not shown in the paper), from the scaling of which we cannot almost distinguish between them. However, we observe that significant changes happen to the amplitude of structure function, which is associated with the possible maximum energy that particle can be reached. In addition, we also studied the relationship between the particle's maximum acceleration rate and the local magnetic field, but no spatial anisotropy of the maximum acceleration rate was found for three modes. We feel that the lack of anisotropic features may be attributed to the limitation of numerical resolution. As pointed out in \cite{Lee16}, the calculation of the structure function, in particular, the recovering of power law scaling, requires a higher numerical resolution; it needs further high resolution numerical simulation in the future.
  
 We would like to say that the mechanism that induces the particles to accelerate is the same, that is, it is a second-order Fermi process for three modes. It is very interesting that although a shock acceleration process is expected for supersonic turbulence, we did not find significant differences compared with subsonic case (see Figure \ref{fig:ptimesubA}). We think that the particle interacting with a single shock leads to the first-order Fermi acceleration, but the first-order effect is canceled out after particles encountering many shock compressions or expansions, and only the second-order Fermi process results in an increase of particle energy. 
 
 From the perspective of numerical practice, the execution of this work consists of three main steps: (1) MHD turbulence simulations to generate data cubes including the information of both magnetic fields and velocities; (2) implementation of wavelet transformation to decompose MHD modes into Alfv\'en, slow and fast modes, and (3) test particle simulations to study the properties of particle acceleration. The cost of simulation time restricts us to use a numerical resolution of 512 along each dimension of coordinates. Especially in the process of test particle simulation, the time-consuming is strongly dependent on the number of injected particles and the terminal time of simulation evolution. The 10,000 test particles we injected have ensured the reliability of the numerical results, which is an order of magnitude higher than the earlier simulation of 1,000 particles (\citealt{Xu13}). As shown in Figure \ref{fig:spec_maxmf}, the inertial range of the MHD turbulence cascade revealed from the particle's maximum acceleration rate is narrower than that displayed by the underlying magnetic field (or velocity). Nevertheless, the execution of higher numerical simulations is necessary to confirm our research results.
 
This work did not include the effect of radiative losses, such as synchrotron emission and inverse Compton scattering, on the particle acceleration. Although the radiative loss is almost negligible for the proton considered as a probe particle in this paper, the loss effect should be included anyway in the some astrophysical environments dominated by hadrons. The radiative loss taken into account, the spectral energy distributions predicted by the current work might vary in the very high-energy regime. Related testing and verification are welcome.   

\section{Summary} \label{Sum}
Considering four turbulence regimes that could appear in the realistic astrophysical settings, we have performed the study of particle acceleration in compressible MHD turbulence using the test particle method. The main results that we found are summarized as follows:

\begin{enumerate}[wide, labelwidth=!, labelindent=1pt]
\item The acceleration of particles caused by Alfv\'en, slow and fast modes is a second-order Fermi process in four turbulence regimes explored in this paper. The transition of particle distributions, from $N\propto t^{3.6}$ to $N\propto t^{0.8}$ in the acceleration process, is related to the injection scale of turbulence, or, to put it another way, it is related to the strong or weak turbulence regimes. 

\item The power spectrum of the particle's maximum acceleration efficiency directly reflects the inertial range of MHD turbulence cascade, and recovers the scaling relations of Alfv\'en, slow and fast modes. In contrast to the inertial scale range revealed by the underlying magnetic fields, the power spectrum of maximum acceleration rate provides a narrower inertial range. Our studies indeed demonstrated that fast mode shows an ideal scaling of $\propto k^{-3/2}$ as expected, and Alfv\'en and slow modes can also recover approximately the scaling of  $\propto k^{-5/3}$.

\item Spectral energy distributions of the accelerated particles for three modes first present a power law form in the low energy range, and then transfers into thermal-like spectrum in the high energy range. 

\item Fast mode dominates the acceleration of particles during the most evolution time, especially in the subsonic turbulence. In the case of sub-Alfv\'en turbulence, slow mode dominates the particle acceleration in the very high energy range. Different from the earlier theoretical predictions, the contribution of Alfv\'en mode cannot be ignored, which even plays a dominant role in the particle acceleration at the very early stage of evolution.  
\end{enumerate}

\acknowledgments
We thank the anonymous referee for valuable suggestions that significantly improved this paper. We would like to thank Grzegorz Kowal for the support of numerical techniques regarding wavelet decomposition and test particle method, Jungyeon Cho for generating data cube of MHD turbulence, and Alex Lazarian for valuable discussions about the particle acceleration in the compressible MHD turbulence. J.F.Z. thanks the supports from the National Natural Science Foundation of China (grant Nos. 11973035 and 11703020) and the Hunan Province Innovation Platform and Talent Plan–HuXiang Youth Talent Project (No. 2020RC3045).  F.Y.X. acknowledges the support from the Joint Research Funds in Astronomy U2031114 under cooperative agreement between the National Natural Science Foundation of China and the Chinese Academy of Sciences.

\end{document}